\documentclass[conference]{IEEEtran}
\IEEEoverridecommandlockouts
\usepackage{cite}
\usepackage{amsmath,amssymb,amsfonts}
\usepackage{algorithmic}
\usepackage{graphicx}
\usepackage{textcomp}
\usepackage{xcolor}
\def\BibTeX{{\rm B\kern-.05em{\sc i\kern-.025em b}\kern-.08em
    T\kern-.1667em\lower.7ex\hbox{E}\kern-.125emX}}
\usepackage{makecell}
\usepackage{amsmath}
\usepackage{bm}
\usepackage{subfig}
\usepackage{booktabs}

\newcommand{\floor}[1]{\left\lfloor #1 \right\rfloor}

\begin{document}

\title{End-to-End JPEG Decoding and Artifacts Suppression Using Heterogeneous Residual Convolutional Neural Network}

\author{\IEEEauthorblockN{1\textsuperscript{st} Jun Niu\textsuperscript{*}}
  \IEEEauthorblockA{
Seattle, WA, USA \\
j.niu1990@gmail.com}
\thanks{\textsuperscript{*}By the time of publishing, Jun Niu is affiliated with Amazon.com, Inc. This work, however, was done prior to the author's joining Amazon.}
\thanks{This is a preprint. To be published by IEEE.}}

\maketitle

\begin{abstract}
Existing deep learning models separate JPEG artifacts suppression from the decoding protocol as independent task. In this work, we take one step forward to design a true end-to-end heterogeneous residual convolutional neural network (HR-CNN) with spectrum decomposition and heterogeneous reconstruction mechanism. Benefitting from the full CNN architecture and GPU acceleration, the proposed model considerably improves the reconstruction efficiency. Numerical experiments show that the overall reconstruction speed reaches to the same magnitude of the standard CPU JPEG decoding protocol, while both decoding and artifacts suppression are completed together. We formulate the JPEG artifacts suppression task as an interactive process of decoding and image detail reconstructions. A heterogeneous, fully convolutional, mechanism is proposed to particularly address the uncorrelated nature of different spectral channels. Directly starting from the JPEG code in k-space, the network first extracts the spectral samples channel by channel, and restores the spectral snapshots with expanded throughput. These intermediate snapshots are then heterogeneously decoded and merged into the pixel space image. A cascaded residual learning segment is designed to further enhance the image details. Experiments verify that the model achieves outstanding performance in JPEG artifacts suppression, while its full convolutional operations and elegant network structure offers higher computational efficiency for practical online usage compared with other deep learning models on this topic.
\end{abstract}

\begin{IEEEkeywords}
JPEG, convolutional neural network, decoding, artifacts suppression
\end{IEEEkeywords}

\section{Introduction}
\label{introduction}
JPEG artifacts suppression aims at removing the noticeable distortion of digital images caused by the lossy compression process. Among the many available image compression protocols, JPEG has earned the dominant popularity due to its high compression ratio and relatively minor humanly perceptible loss in image quality. The high portability and relatively decent quality make JPEG images ideal for transmitting through internet services with limited bandwidth and archiving with affordable storage resources. However, while numerous images have been exchanged and stored in JPEG format, very few of the original lossless raw data remains available. For the applications where users would like to obtain a high quality copy of an accessible JPEG image, removing the compression artifacts and reconstructing the details becomes the most applicable, if not the only, approach. On the other hand, at the age of huge volume of digital images being shared and transmitted through the internet everyday, JPEG protocol's high compression ratio makes it a preferred format to accommodate the limited internet resources. Compared with directly sending the lossless images, first receiving the digital images in JPEG format and then removing the compression artifacts at the users' end serves as a more economic and realistic option.

These specific engineering backgrounds emphasize not only the statistical performance in artifacts suppression, but also the computational efficiency in operation. A few numerical frameworks have been proposed to serve this purpose during the past decades. One typical category of solutions is based on the conventional optimization techniques. For example, Nguyen et al. propose to encode a 64-KB overhead into the JPEG code \cite{Nguyen_2016_CVPR}. The raw image is then reconstructed from the JPEG image together with the overhead by solving the constrained optimization problem. The reconstructed image shows a relatively low error. However, most constrained optimization employs an iterative solver. Since their computational costs are heavy, these approaches hardly qualify for intensive online applications.

Another category of solutions focuses on deep learning models. Although the end-to-end neural network design is a non-trivial work and the training processes are usually expensive. Once the training is completed, however, little pre-/ post-processing would be needed, and prediction on new inputs is purely a feed-forward process. These merits make the deep neural network a promising solution for both offline and online applications. Recently, several deep neural networks have been proposed for image super-resolution \cite{doi:10.1063/1.4993002,7115171} and image denoising \cite{7839189}.  Chen et al. proposes a strategy for compression artifacts suppression inspired by the super-resolution neural networks \cite{CHEN2018204}. Although this is an intriguing research direction, artifacts suppression is not exactly the same problem as image super-resolution. Namely, super-resolution solvers' primary responsibility is to predict the information beyond the sampling band, while the artifacts suppression emphasizes reconstructing information both within and beyond the sampling band. A few other deep convolutional neural networks have also been proposed for JPEG artifacts suppression \cite{7965927,Maleki_2018_CVPR_Workshops,7410430,Svoboda2016}. However, they usually have complex structure with few domain-specific insights. The fundamental characters of the JPEG protocol is not fully analyzed to optimize the network structural design. The corresponding complex architecture leads to unnecessarily large amount of model parameters and extremely expensive training process.  More importantly, encoding and decoding are central problems in communication \cite{MacKay_2002}. Existing designs split decoding and artifacts suppression as two independent steps. For most main stream compression protocols, such as JPEG, this strategy is not end-to-end. A decoder must be first called to transform JPEG code from k-space to pixel space, and artifacts are then removed from compressed images in pixel domain. The corresponding computational cost adds additional burden for real-time applications.

In this work, we propose a novel heterogeneous residual convolutional neural network (HR-CNN) for JPEG compression artifacts suppression. The proposed HR-CNN combines decoding and artifacts suppression into one end-to-end fully convolutional network, where all segments are optimized interactively. The relatively simplified structure, in turn, also delivers considerably higher efficiency in operation. Directly loading the k-space JPEG code into the input layer, the neural network creates the optimized image in pixel space as final output. For practical online usage, the network achieves fast processing, and is less memory consuming due to its simplified network structure and fully convolutional operations. To address the uncorrelated nature of k-space codes in different spectral channels, a spectral decomposition mechanism and heterogeneous convolutional operation pipeline is designed to complete the decoding process. Connecting the created intermediate features to a residual learning segment, the decoding and detail reconstruction process are then interactively optimized. Numerical experiments demonstrate that the proposed design shows excellent performance in the peak signal-to-noise ratio (PSNR) and structural similarity index (SSIM). Benefitting from the full CNN architecture and GPU acceleration, it completes the the end-to-end decoding and artifacts suppression within comparable time of image decoding following the standard CPU JPEG protocol. The overall operational efficiency is far superior to existing two-step models.

Overall, the contributions of this study are mainly in three aspects:
\begin{enumerate}
    \item To our best knowledge, it is the first true end-to-end model designed for JPEG decoding and artifacts suppression. Directly loading the k-space JPEG code into the input layer, a pixel space image is reconstructed as the model's output.
    \item A novel heterogeneous and fully convolutional mechanism is proposed to address the uncorrelated nature of spectral channels. Versatile engineerings are thus accommodated for features in different spectral channels. 
    \item The proposed design delivers outstanding performance under specific training criteria. In the meantime, its simplified network structure largely improves the computational efficiency in practical online usage.
\end{enumerate}

\section{Heterogeneous Residual Convolutional Neural Network for JPEG Artifacts Suppression}
\label{model design}
\subsection{Design Overview}

The proposed neural network for JPEG artifacts suppression focuses on two key concepts: heterogeneous processing across spectral channels and true end-to-end reconstruction. The standard JPEG protocol encodes digital images with 64 uncorrelated spectral channels. The heterogeneous processing principle aims to offer a flexible mechanism to engineer the features in each channel according to their statistical characteristics. Designing the network as true end-to-end avoids additional computational burden caused by preprocessing, thus enhances its candidacy for intensive online usage.

\begin{figure}[ht]
\begin{center}
\centerline{\includegraphics[width=0.51\textwidth]{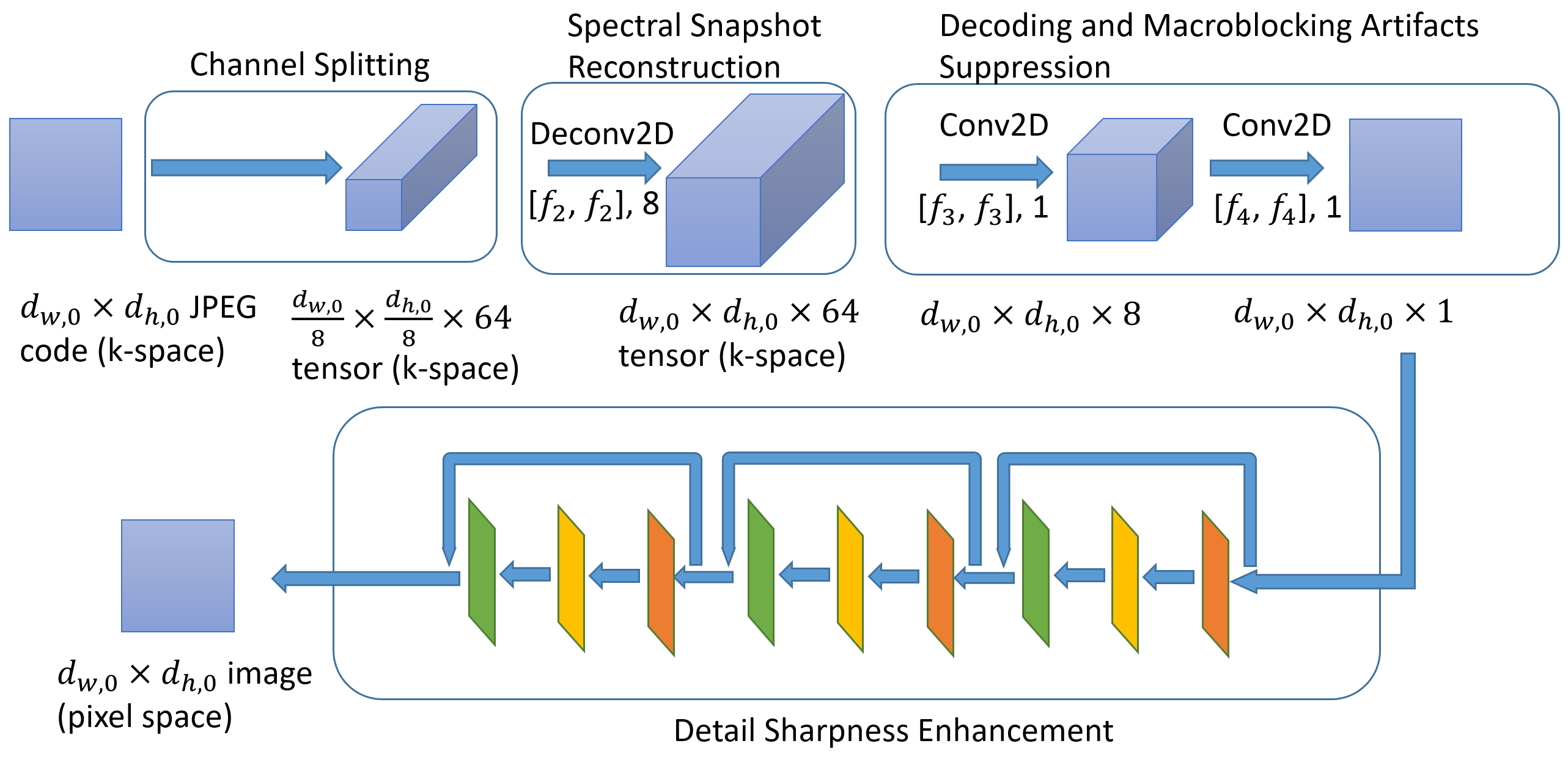}}
\caption{Structural illustration of the heterogeneous residual convolutional neural network.}
\label{nn structure and overview}
\end{center}
\end{figure}

Figure \ref{nn structure and overview} provides an overview of the processing pipeline and the network structure. An end-to-end JPEG artifacts suppression model maps the k-space JPEG code to a pixel-space image as close to the ground truth as possible. Directly loading the JPEG code in k-space into the input layer, the proposed model first decomposes the JPEG code into 64 spectral channels through parallel convolutional operations. Here the convolution kernels are designed as a set of coded masks. For each spectral channel, a full-size snapshot is created through the one-to-many transposed convolutional mapping. The obtained spectral space features are then fed into a decoding module to obtain pixel space counterpart. The heterogeneous nature of this conceptual decoding process enforces flexible operations on different channels. A residual learning segment then further reconstructs the details of the pixel space image and generates the final output. All network segments are optimized interactively at the training process.

\subsection{Formulation}
\label{section: formulation}
Consider a k-space image encoded from the original pixel space image, $\mathbf{X}$, through the standard JPEG compression protocol. Denoting this k-space image as $\mathbf{Y}$, our goal is to recover an image $F(\mathbf{Y})$ that is as close as possible to the ground truth image $\mathbf{X}$. The proposed neural network wishes to learn a powerful mapping $F(\cdot)$, which conceptually achieves four operations: channel extraction and representation, spectral snapshot reconstruction, decoding and macroblocking artifacts suppression, and image detail sharpness enhancement.

\subsubsection{Channel extraction and representation}
\label{section: channel extraction}
\begin{figure}[ht]
\begin{center}
\centerline{\includegraphics[width=1.0\columnwidth]{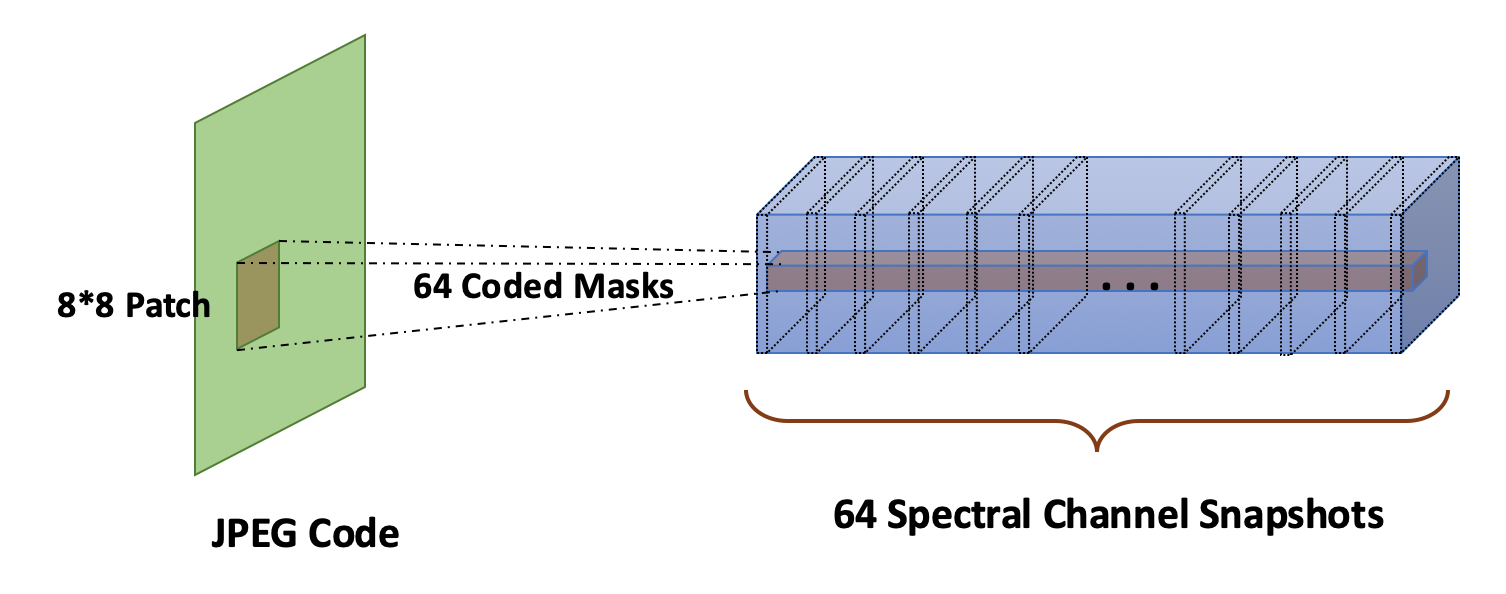}}
\caption{Convolutional operations to extract samplings in each channel as one spectral space snapshot.}
\label{channel splitting}
\end{center}
\end{figure}

\begin{figure}[ht]
\begin{center}
\centerline{\includegraphics[width=1.0\columnwidth]{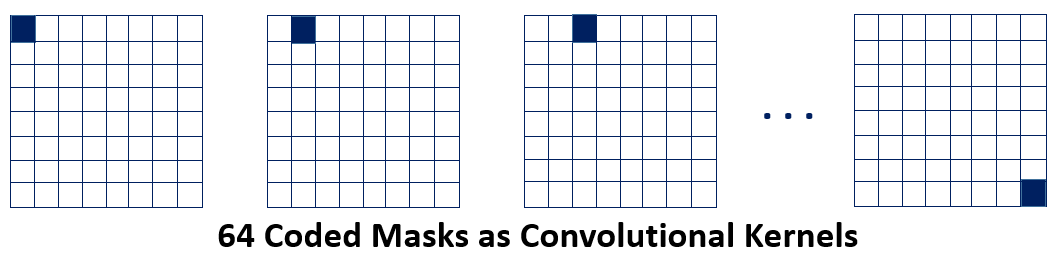}}
\caption{Convolution kernels designed as coded masks. The dark blue blocks denote the matrix entries with value 1; the white blocks denote the matrix entries with value 0.}
\label{coded masks}
\end{center}
\end{figure}

The k-space code $\mathbf{X}$ consists of the samples across 64 spectral channels, grouped by $8 \times 8$ macro blocks. This network segment extracts the spectral samples in $\mathbf{X}$ channel by channel, and represents them as a set of spectral space snapshots. Fig. \ref{channel splitting} provides a sketch of this process. Channel extraction serves as the foundation for the subsequent heterogeneous operation across different spectral channels. A special convolutional layer is designed to achieve this goal efficiently. In contrast with the standard convolutional layers, the convolution kernels in this layer are engineered as 64 binary coded masks. As is illustrated in Fig. \ref{coded masks}, each binary coded mask is initialized with 63 zero-value entries and 1 one-value entry. In operation, each mask extracts the snapshot at its assigned channel and maps the $d_{w,0} \times d_{h,0}$ dimensional input $\mathbf{X}$ into a $\frac{d_{w,0}}{8} \times \frac{d_{h,0}}{8}  \times 64 $ tensor.

Formally, the first layer is expressed as an operation $F_1$:
\begin{equation}
F_1(\mathbf{Y}) = \mathbf{W_1} * \mathbf{Y} + \mathbf{B_1}
\end{equation}
where $\mathbf{Y}$ is the input JPEG code;  $*$ denotes the convolution operator; $\mathbf{W_1}$  contains $n_1$ filters of size $c \times f_{1x} \times f_{1y}$ for channel extraction and scaling, with $c$ denoting the number of spectral channels of the input JPEG image code, and $f_{1x}$, $f_{1y}$ denote the number of orthogonal spectral components in $x$ and $y$ directions, respectively. Within the context of JPEG artifacts compression, we have $f_{1x} = 8$ and $f_{1y} = 8$. The stride in $x$ direction is correspondingly set identical to $f_{1x}$ and the stride in $y$ direction is set identical to $f_{1y}$. $\mathbf{B_1}$ is $c-$dimensional, which realizes the translation operation for each channel.

To extract the spectral components channel by channel, each convolution kernel is designed to maintain the information carried at one assigned channel, denoted as $ch$, while the information in all remaining channels is shielded from both network training and feed-forward computation:
\begin{equation}
M_{ij}^{ch}  = \delta(i - mod(ch, 8), j - \floor{ch / 8}), i, j = 0, 1, ..., 7.
\end{equation}
Here $\delta$ denotes the Dirac delta function.

The coded masks at each channel is a pre-designed constant tensor. At the back propagation of the training process,
\begin{equation}
\Delta F_1^{ch} = \mathbf{M}^{ch} * \Delta \mathbf{Y_1} 
\end{equation}
The gradient of channel $ch$ selected by none-zero component of $\mathbf{M}^{ch}$ propagates freely across the network. On the contrary, gradients of all other channels are reset to 0 through the zero components of $\mathbf{M}^{ch}$, thus have no influence on the training of any other layer.

The subsequent network layers will take the responsibility to enforce the convolutional operations for decoding and artifacts suppression. Channel-specific value translation is not necessary at this step. As a result, we set the bias term $\mathbf{B_1}$ is set to be $\mathbf{0}$. 

In summary, the first layer extracts the snapshot for each spectral channel through the following mapping,
\begin{equation}
\label{eq: channel splitting}
F_1^{ch}(\mathbf{Y_1}) = \mathbf{M}^{ch}  * \mathbf{Y_1} , ch = 0, 1, ..., 63.
\end{equation}

\subsubsection{Spectral Space Reconstruction}
\label{section: spectral domain reconstruction}
The snapshots at each spectral channel are reconstructed to the same dimension as that of the original image in this layer. One of the major sources of macroblocking artifacts comes from the quantization step of JPEG compression. The magnitudes of high frequency components are stored with reduced resolution. To address this issue, we tackled the challenging task of enriching the throughput in each layer. The JPEG compression protocol enforces a many-to-one mapping for the components in each of the 64 channels. As a results, each snapshot extracted through Eq. \ref{eq: channel splitting} is of dimension $\frac{d_{w,0}}{8} \times \frac{d_{h,0}}{8}$. Before moving into the decoding and artifacts suppression operations, we first reconstruct each snapshot to its original size through the transposed convolutional operation. This step conceptually performs an inverse operation of compression and restores the one-to-many positional connectivity:

\begin{equation}
\label{eq: transposed conv layer}
\tilde{F}_2(\mathbf{Y}) = \mathbf{C}^T_2 \cdot \tilde{F}_1(\mathbf{Y}) + B_2
\end{equation}

where $\cdot$ denotes the matrix multiplication; $\mathbf{C}_2$ is the matrix operator of the convolutional operation with kernel $\mathbf{W_2}$; $\tilde{F}_1(\cdot)$ and $\tilde{F}_2(\cdot)$ denote the row-major order vector representation of the tensor outputs of $F_1(\cdot)$ and $F_2(\cdot)$, respectively. The convolution kernel $\mathbf{W_2}$ contains $c$ filters of size $c\times f_2 \times f_2$, where $c$ denotes the number of channels in the k-space JPEG code; $f_2$ is chosen as the block size, 8, so that each snapshot will be enriched to the full image's size. The bias term $B_2$ is $c$-dimensional. The total number of parameters in this layer is exactly the same as that in the standard convolutional layer. the computational cost in training and prediction process thus will not be significantly impacted.

Through this layer, the dimension of intermediate features will be mapped from $64 \times \frac{d_{w,0}}{8} \times \frac{d_{h,0}}{8}  $ to $64 \times d_{w,0} \times d_{h,0} $.

\subsubsection{Decoding and Macroblocking Artifacts Suppression}
Once the spectral space samplings have been tentatively enriched, we continue to the decoding process. The standard JPEG decoding protocol directly applies the inverse discrete cosine transform to each $8\times8$ block, through which each pixel is constructed with the spectral samplings on the same row and the same column. This process is robust due to the uncorrelated nature of different spectral channels. On the contrary, however, the enriched snapshots from Eq. \ref{eq: transposed conv layer} is obtained from the transposed convolutional one-to-many mapping. The decoding process is correspondingly modified to be a 3-dimensional convolutional operation. To offer more flexible processing capacity, a network segment consisting of 2 cascading convolutional layers is designed to realize the decoding process. The macroblocking artifacts are also expected to be partly suppressed with the increased spectral throughput.

At each step $i$, the convolution layer perform the feature engineering as
\begin{equation}
F_i(\mathbf{Y}) = \mathbf{W_i} * F_{i-1}(\mathbf{Y}) + B_i, i = 3, 4.
\end{equation}
Here $\mathbf{W_3}$ corresponds to $n_3$ filters of size $c \times f_3 \times f_3$; $B_3$ is $n_3$-dimensional, with $n_3 = 8$, $f_3 = 5$. $\mathbf{W_4}$ corresponds to $n_4$ filters of size $n_3 \times f_4 \times f_4$; $B_4$ is $n_4$-dimensional, with $n_4 = 1$, $f_4 = 3$. For all layers in this segment, the convolutional strides are chosen as 1 in both $x$ and $y$ directions. A ReLu activation is applied to $F_5(\mathbf{Y}) $ to futher enforce the non-negative regularization.

This network segment conceptually constructs a pixel space image of the original dimension $d_{w,0} \times d_{h,0}$ from the $64 \times d_{w,0} \times d_{h,0} $ dimensional spectral space features.

\subsubsection{Image Detail Sharpness Enhancement}
\label{section: residual learning}
After the initial suppression of macroblocking artifacts in pixel space, extra reconstruction of the details is necessary to improve the overall image sharpness. One potential approach to restore the image details from the lossy compression is to build a mapping directly from the crude intermediate features with lower sharpness and resolution. However, it turns out to be very difficult to train a very deep convolutional neural network with sufficient emphasis on these details. One of the fundamental reasons is that for most typical choices of loss functions, such as the mean squared error, the errors caused by the image details get completely overwhelmed by that caused by the macroblocking artifacts. As an alternative, here we apply the cascaded residual learning blocks design to make the neural network optimization significantly easier. The key principle of each residual learning block is to decompose the higher-resolution image details from the lower-resolution image bulks. Convolutional layers inside each residual learning block conceptually focus on the processing of higher-resolutions details. After merging with the lower-resolution image bulks, features of the complete image are delivered to the next cascaded residual learning block for further enhancement.

Operations in this network segment can be formally expressed as:
\begin{equation}
\begin{split}
F_i(\mathbf{Y}) = \mathbf{W_i} * F_{i-1}(\mathbf{Y}) + B_i + F_{i-3} (\mathbf{Y}) \cdot [\delta(i - 7) +  \\
\delta(i - 10) + \delta(i - 13) + \delta(i - 16)], i = 5, 6, ..., 16.
\end{split}
\end{equation}

In the $i$-th convolutional layer, $\mathbf{W_i}$ contains $n_i$ filters of dimension $n_{i-1} \times f_i \times f_i$ and $B_i$ is $n_i$-dimensional. The convolutional stride is taken as 1 in both $x$ and $y$ directions. In the first residual block, we choose $n_5$, $f_5$ as $64$ and $11$, respectively; $n_6$, $f_6$ as $16$ and $7$, respectively; $n_7$, $f_7$ as $1$ and $1$, respectively. The same design pattern repeats for the next three residual blocks, which consist of layers $i = 8, 9, 10$,  $i = 11, 12, 13$ and $i = 14, 15, 16$, respectively. 

The direct connection between the input and output of each residual learning block largely increases the training feasibility. In each block, the three convolutional layers optimize the image details through a data-driven manner. The layer with unit-size convolutional kernels aims to improve the smoothness of the restored image.

After completing all the operations to enhance the image detail sharpness, the final output is reconstructed through a ReLu activation:
\begin{equation}
F(\mathbf{Y}) = ReLu(F_{16}(\mathbf{Y}) ).
\end{equation}
This step not only enforces non-negative regularization to the pixel value, but also renders the neural network nonlinear processing capability. The network reconstructs the pixel space image of the original size, $d_{w,0} \times d_{h,0}$.

\subsection{Training}
Learning the end-to-end mapping function $F(\cdot)$ requires the estimation of the network parameters $ \Theta = \{\mathbf{W_i}, B_i \}, i = 1, 2, ..., 16$. Given a set of ground truth lossless images $\{\mathbf{X_i} \}$ and the corresponding k-space codes $\{\mathbf{Y_i} \}$ created through standard JPEG compression protocol, we use the mean squared error (MSE) as the loss function:

\begin{equation}
\label{eq: loss function}
L(\mathbf{\Theta}) = \frac{1}{n} \sum_{i=1}^{n} || F(\mathbf{Y_i}; \Theta) - \mathbf{X_i}||^2 ,
\end{equation}

where $F(\mathbf{Y_i}; \Theta)$ is neural network's output corresponds to $\mathbf{X_i}$; $n$ is the number of training image pairs. Learning of the network parameters $\Theta$ is then completed through the minimizing the pixel level difference between the ground truth and the reconstructed image.

Minimizing Eq. \ref{eq: loss function} through the stochastic gradient descent algorithm with back-propagation, all network segments analyzed in Section \ref{section: formulation} are optimized interactively. Choosing the MSE as the loss function, the trained neural network equivalently optimizes the reconstructed images' peak signal-to-noise ratio (PSNR). However, according to specific engineering applications, the loss function can be easily revised to address other perceptual quality metrics \cite{Johnson2016PerceptualLF}.

\section{Experiments and Performance Evaluation}
\label{experiments and performance evaluation}

\begin{figure*}[ht]
\begin{center}
\captionsetup[subfigure]{justification=centering}
\subfloat[][Model input - JPEG k-space code.]{\includegraphics[width=0.33\textwidth]{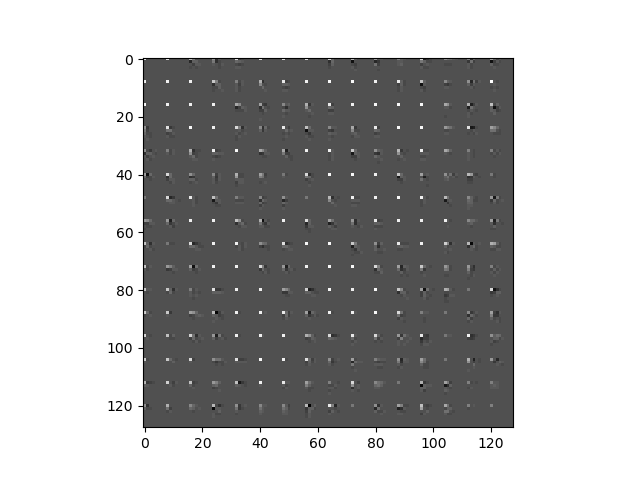}} 
\subfloat[][1st channel snapshot after channel-splitting.]{\includegraphics[width=0.33\textwidth]{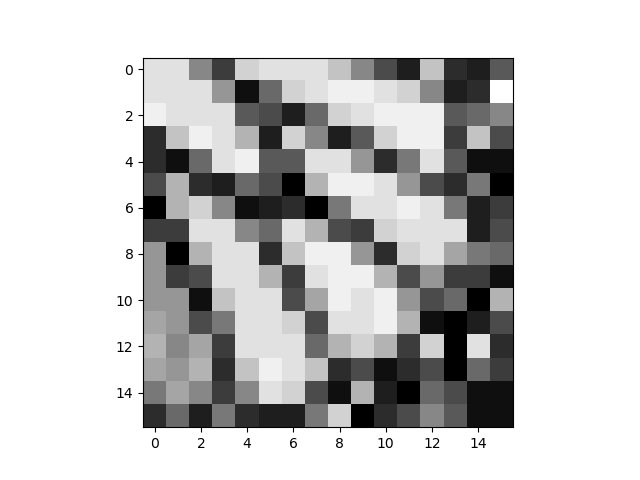}}
\subfloat[][1st channel snapshot after spectral-reconstruction]{\includegraphics[width=0.33\textwidth]{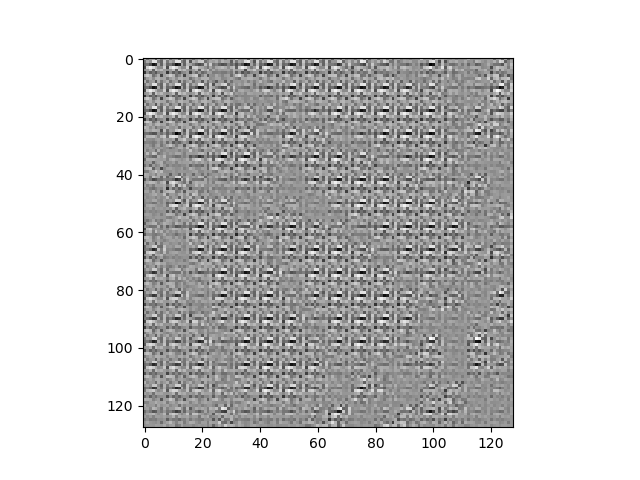}} 
\\[-2ex]
\noindent 
\subfloat[][Snapshot after decoding.]{\includegraphics[width=0.33\textwidth]{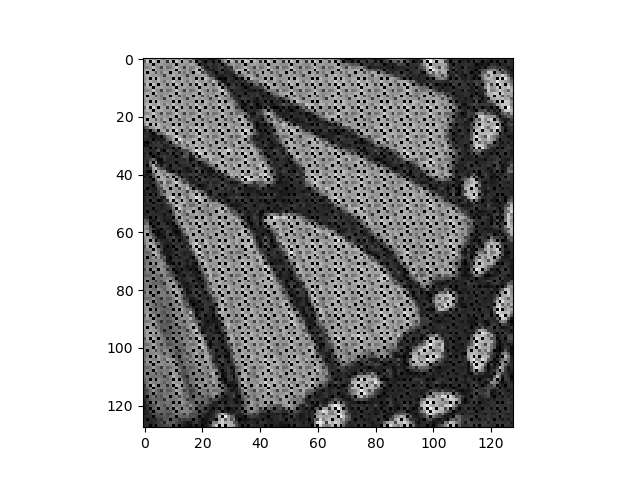}} 
\subfloat[][Snapshot after the 2nd residual block in sharpness-enhancement.]{\includegraphics[width=0.33\textwidth]{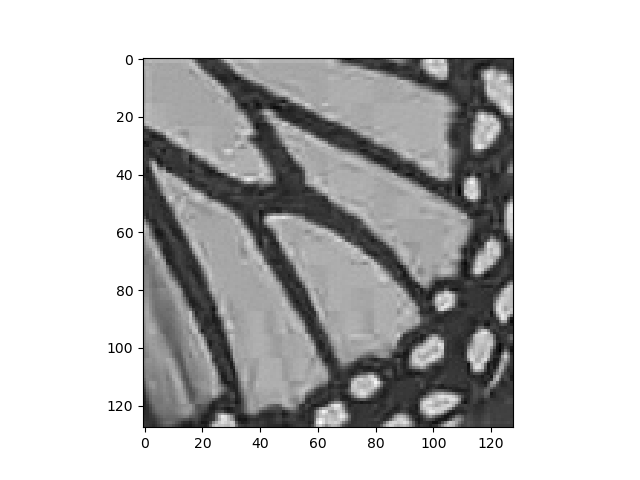}}
\subfloat[][Model output - after the 4th residual block in sharpness-enhancement.]{\includegraphics[width=0.33\textwidth]{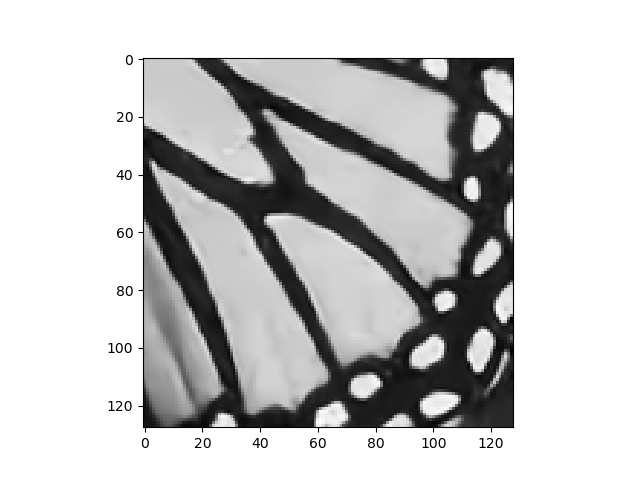}} 
\end{center}
\caption[caption]{Ablation Study at Q = 10.} 
\label{ablation study}
\end{figure*}

\begin{figure*}[ht]
\begin{center}
\captionsetup[subfigure]{justification=centering}
\subfloat[][Image A, Q = 10, JPEG, \\ PSNR =  25.16 dB; SSIM = 90.33\%]{\includegraphics[width=0.33\textwidth]{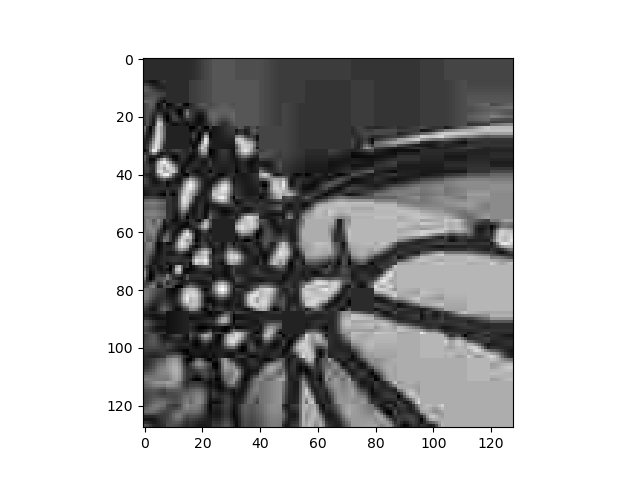}} 
\subfloat[][Image A, Q = 10, Reconstructed, \\ PSNR =  28.20 dB; SSIM = 95.12\%]{\includegraphics[width=0.33\textwidth]{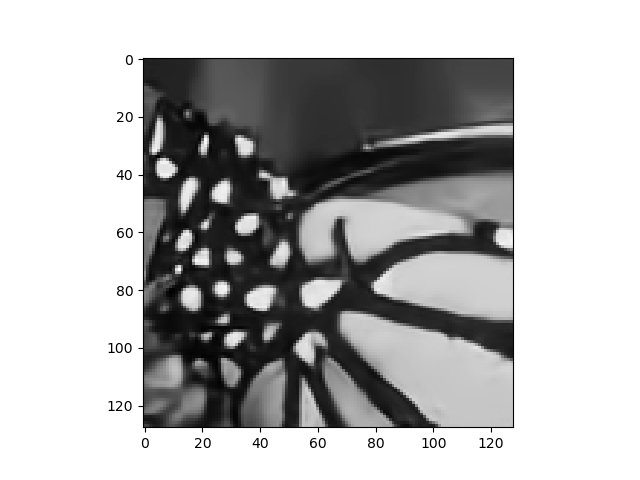}}
\subfloat[][Image A, Q = 30, JPEG, \\ PSNR =  28.94 dB; SSIM = 95.56\%]{\includegraphics[width=0.33\textwidth]{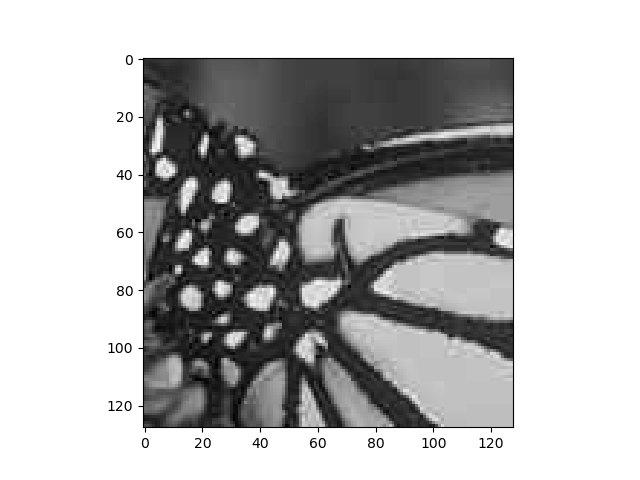}} 
\\[-2ex]
\noindent 
\subfloat[][Image A, Q = 30, Reconstructed, \\ PSNR =  31.93 dB; SSIM = 97.73\%]{\includegraphics[width=0.33\textwidth]{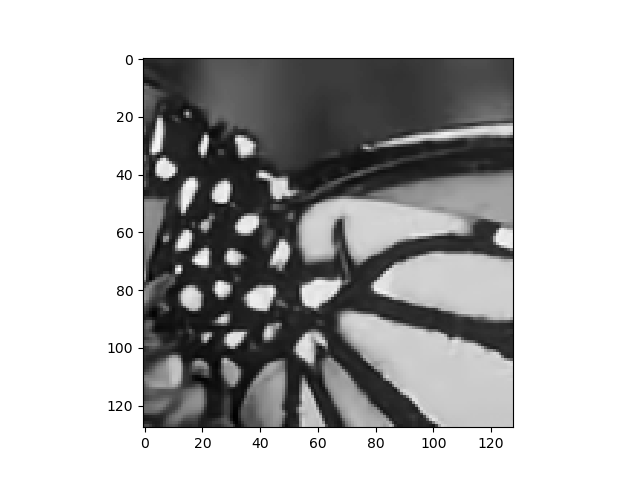}} 
\subfloat[][Image A, Q = 50, JPEG, \\ PSNR =  30.77 dB; SSIM = 96.93\%]{\includegraphics[width=0.33\textwidth]{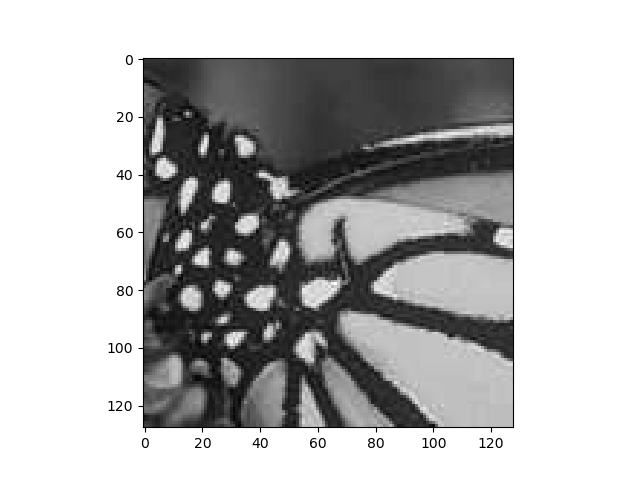}}
\subfloat[][Image A, Q = 50, Reconstructed, \\ PSNR =  33.87 dB; SSIM = 98.39\%]{\includegraphics[width=0.33\textwidth]{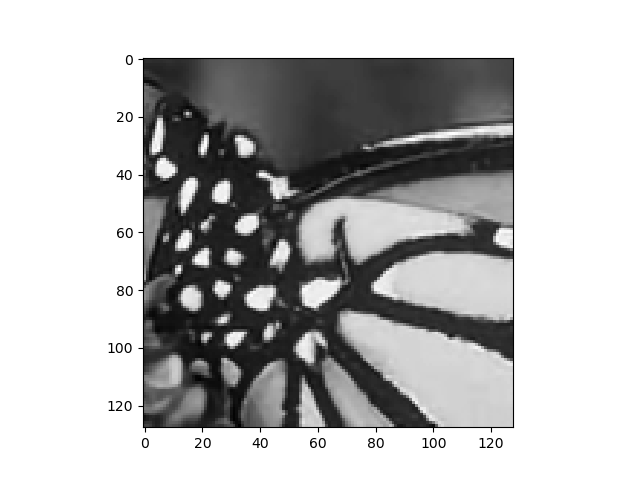}} 
\\[-2ex]
\captionsetup[subfigure]{justification=centering}
\subfloat[][Image B, Q = 10, JPEG, \\ PSNR =  23.63 dB; SSIM = 87.50\%]{\includegraphics[width=0.33\textwidth]{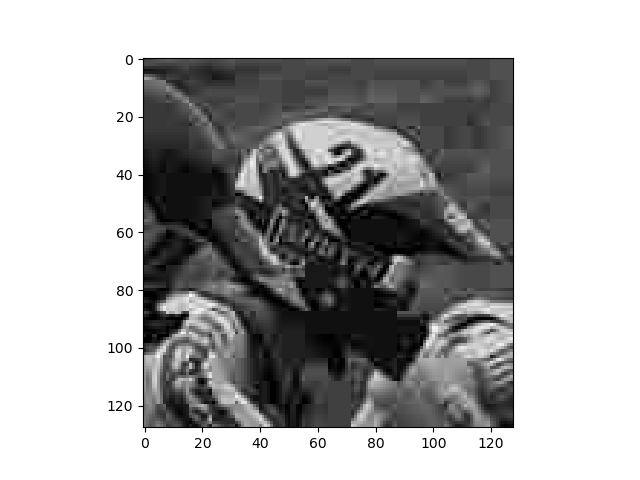}} 
\subfloat[][Image B, Q = 10, Reconstructed, \\ PSNR =  26.10 dB; SSIM = 92.34\%]{\includegraphics[width=0.33\textwidth]{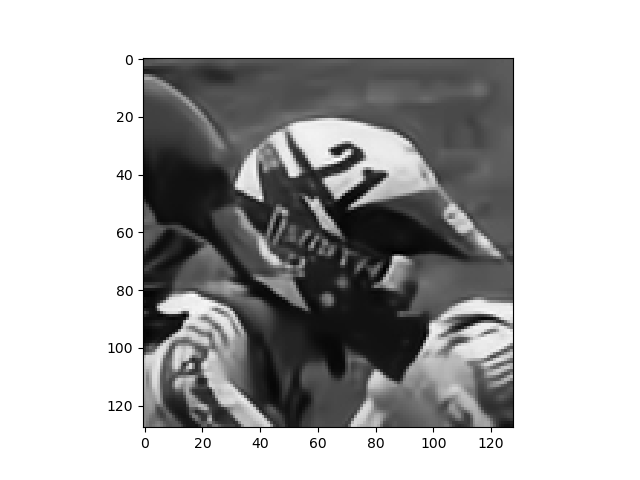}}
\subfloat[][Image B, Q = 30, JPEG, \\ PSNR =  27.39 dB; SSIM = 94.60\%]{\includegraphics[width=0.33\textwidth]{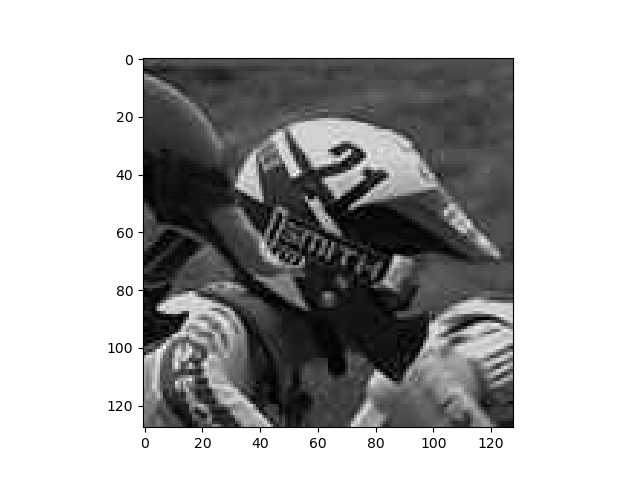}} 
\\[-2ex]
\noindent 
\subfloat[][Image B, Q = 30, Reconstructed, \\ PSNR =  30.41 dB; SSIM = 97.14\%]{\includegraphics[width=0.33\textwidth]{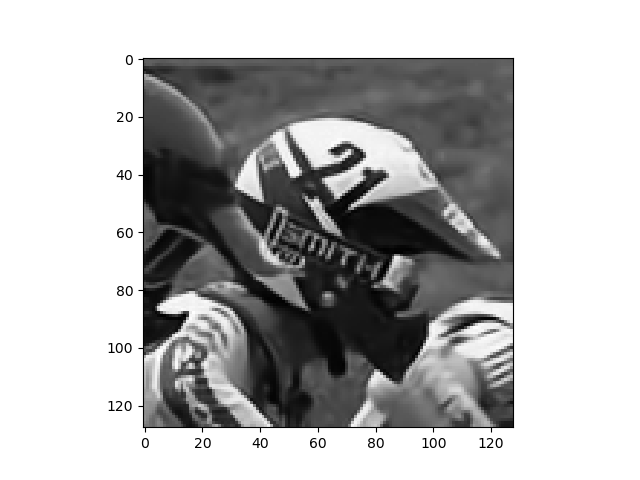}} 
\subfloat[][Image B, Q = 50, JPEG, \\ PSNR =  29.48 dB; SSIM = 96.41\%]{\includegraphics[width=0.33\textwidth]{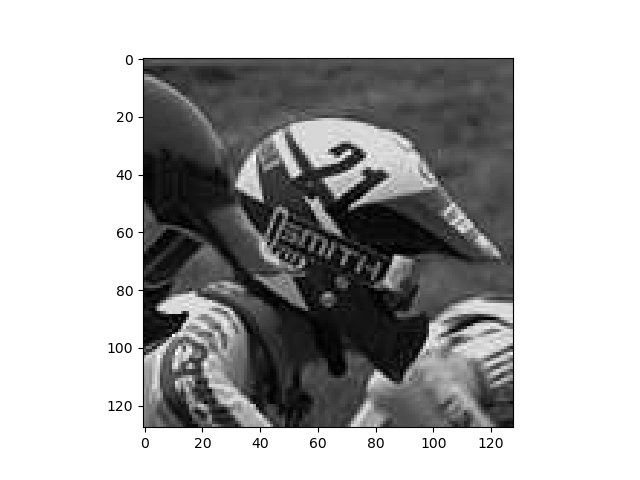}}
\subfloat[][Image B, Q = 50, Reconstructed, \\ PSNR =  32.42 dB; SSIM = 98.14\%]{\includegraphics[width=0.33\textwidth]{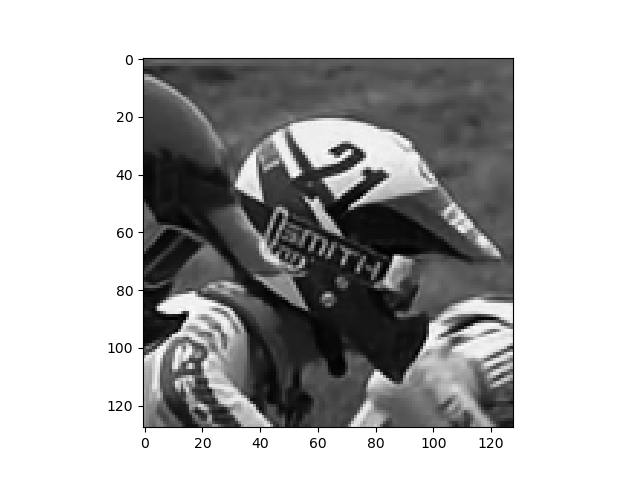}} 
\end{center}
\caption[caption]{Comparison of Two Typical Sets of JPEG Images and Reconstructed Images at Q = 10, 30, 50.} 
\label{comparison figure 32 - 3}
\end{figure*}

\subsection{Training Datasets}
We crop the training, validation, and testing datasets from the DIV2K dataset \cite{Agustsson_2017_CVPR_Workshops,Timofte_2017_CVPR_Workshops,Timofte_2018_CVPR_Workshops}. The DIV2K dataset consists of 1,000 high quality images of 2K resolution. All images are originally stored in the PNG format. We consider these images and any sub-images directly cropped from them as the ground truth that contains no JPEG compression artifacts. Without loss of generality, we set the dimension of training images as $128 \times 128$, though the proposed model is insensitive to the input image size. Generating  $\mathbf{Y}$ from  $\mathbf{X}$ follows the standard JPEG compression protocol, and is the only processing we perform to create training, validation, and testing datasets. For each $Q$ factor, the training dataset consists of 125,000 images at $128 \times 128$ resolution. Once the model is trained, it can reconstruct images at arbitary resolution.

\subsection{Experiments}

Based on the ground truth lossless images cropped from the DIV2K dataset, we follow the standard JPEG compression process to create the k-space JPEG codes \cite{DBLP:journals/corr/RaidKEA14}. Without loss of generality, we only extract the green channel of the RGB images for further $128 \times 128$ resolution image cropping. The discrete cosine transform is then applied to each $8 \times 8$ block of the obtain images and transform the image from pixel space to spectral space. Element-wisely dividing each $8 \times 8$ block with the quantization matrix and rounding the floating point values, the k-space JPEG code is obtained from the ground truth counterpart. Feeding these k-space JPEG codes into the heterogeneous convolutional neural network, the network's outputs are directly compared with the ground truth images for training loss calculation or model performance evaluation.

Without loss of generality, here we evaluate the proposed neural network's performance at three typical quantization factors: $Q = 10, 30, 50$. The network can be trained at any other $Q$ factors. However, at $Q = 50$, the perceptual quality of the JPEG images is already high. The actual needs for artifacts suppression at even higher quantization factors are not as critical in practical applications. According to our experience, the whole training process usually completes in less than 12 hours on single NVIDIA 2080 GPU.

A detailed ablation study is completed to validate the model's design strategy. During prediction, intermediate features are extracted after channel-splitting, spectral-reconstruction, decoding, and sharpness-enhancement, respectively. Without loss of generality, Fig. \ref{ablation study} provides one typical set of results. The evolvement of the feature snapshots aligns with the expected functionality of each network segment and thus successfully verifies the network design.

The standard JPEG decoding protocol is one special implementation of $F(\mathbf{Y})$. As the lossy nature of JPEG compression, the k-space JPEG code stores the high frequency components with reduced resolution and discards part of the spectral space information. The standard JPEG decoding mechanism cannot, and is not expected to, compensate the information lost during the compression process. Here we implemented a decoder following the standard JPEG protocol on CPU, and use it as the baseline benchmark for any JPEG decoding algorithms. The effectiveness of the proposed HR-CNN can be measured through the improvement over the implemented JPEG decoder.

Figure \ref{comparison figure 32 - 3} shows two typical sets of reconstruction results. Although the training loss function is chosen to emphasize the PSNR, we also observe a satisfying increase in the SSIM.

\subsection{Model Performance}
\begin{table}[t]
\caption{Model's Average Performance over 14,040 Images. }
\label{table: model average performance over 14040 images}
\vskip 0.15in
\begin{center}
\begin{tabular}{lcccc}
\toprule
Image Set & \makecell{PSNR \\(dB)} & \makecell{SSIM \\(\%)} & \makecell{IPSNR \\(dB)} & \makecell{ISSIM \\(\%)} \\
\midrule
Q = 10, JPEG    &  30.13 &  89.57  & -- & --  \\
Q = 10, HRCNN    &  31.79 &  92.02 & 1.66  & 2.45 \\
\midrule
Q = 30, JPEG    & 34.29  & 95.10  & -- & --   \\
Q = 30, HRCNN    & 35.70  & 96.13 & 1.41   & 1.03    \\
\midrule
Q = 50, JPEG    & 36.06  & 96.52 & -- & --  \\
Q = 50, HRCNN    & 37.19  & 97.19  & 1.13  & 0.67    \\
\bottomrule
\end{tabular}
\end{center}
\vskip -0.1in
\end{table}

\begin{table}[t]
\caption{Model's Performance on LIVE1 Dataset. }
\label{table: model performance}
\vskip 0.15in
\begin{center}
\begin{tabular}{lcccc}
\toprule

Image Set & \makecell{PSNR \\(dB)} & \makecell{SSIM \\(\%)} & \makecell{Time Cost \\(s)} \\
\midrule
Q = 10, JPEG    &  27.19 &  87.32 &  2.81  \\
Q = 10, HR-CNN    &  \bf{28.64} &  \bf{89.82} & 6.29  \\
\midrule
Q = 30, JPEG    & 31.00  & 94.21  & 2.84  \\
Q = 30, HR-CNN    & \bf{32.41}  & \bf{95.37}  & 6.27 \\
\midrule
Q = 50, JPEG    & 32.87 & 96.03  & 2.79 \\
Q = 50, HR-CNN    & \bf{34.15}  & \bf{96.82}  & 6.31 \\
\bottomrule
\end{tabular}
\end{center}
\vskip -0.1in
\end{table}

\begin{table}[t]
\caption{Model Comparison on LIVE1 Dataset. }
\label{table: model comparison}
\vskip 0.15in
\begin{center}
\begin{tabular}{c|ccc}
\toprule

Model &  \makecell{IPSNR \\(dB)} & \makecell{ISSIM\\ (\%)} & \makecell{Number of \\Trainable Parameters}  \\
\midrule
\makecell{HR-CNN}    & \bf{1.44}  & \bf{2.50} &  \makecell{Decoding: \\275,089 \\ Detail Enhancement: \\232,068 \\ Total: \\\bf{507,157}} \\
\midrule
\makecell{AR-CNN  \cite{7410430} }    & 1.19   & 3.1  & 106,448   \\
\midrule
\makecell{SA-DCT  \cite{4154791} }    & 0.88   & 1.88  & --   \\
\midrule
\makecell{L4  \cite{Svoboda2016} }    & 1.31   & 3.3  & 71,920   \\
\midrule
\makecell{MemNet \cite{memnet-a-persistent-memory-network-for-image-restoration} }   & 1.68   & 4.6  & 2,905,421   \\
\midrule
\makecell{TNRD \cite{7527621}  }  & 1.38   & 3.8  & 26,645   \\
\midrule
\makecell{DnCNN \cite{7839189}  }  & 1.42   & 3.9  & 668,225   \\
\midrule
\makecell{D-GAN \cite{Galteri2017DeepGA} }   & -0.48   & -1.8  & --   \\
\midrule
\makecell{DA-CAR  \cite{Albluwi2018} }   & 1.31   & 3.2  & 106,336   \\
\midrule
\makecell{CAS-CNN \cite{7965927} }   & 1.67   & 4.2  & 5,144,000   \\
\midrule
\makecell{SA-CAR  \cite{Albluwi2018} }   & 1.39   & 3.5  & 131,392   \\
\bottomrule
\end{tabular}
\end{center}
\vskip -0.1in
\end{table}

The model's statistical performance is evaluated over two datasets: (1) a testing dataset of 14,040 images at  $128 \times 128$ resolution, each cropped from the testing images in DIV2K dataset; (2) LIVE1 dataset \cite{1709988,1284395}. None of the images in both datasets are involved in any model training/ hyper-parameter tuning process. The relatively large number of images in the first testing dataset provides an inspection of the model generalization capability in production. Meanwhile, testing on LIVE1 dataset aims to provide a performance comparison with existing artifacts suppression models.

Here we interpret the image reconstruction quality through two perspectives: the meaningful signal power in the reconstructed image over the noise (PSNR), and the perceptual structure quality of the reconstructed image (SSIM). Table \ref{table: model average performance over 14040 images} shows the network's average performance over 14,040 images in the first dataset. The training process renders the neural network a preference on optimizing the PSNR by using the MSE as the loss function. Nevertheless, significant improvements are observed for both PSNR and SSIM. 

Table \ref{table: model performance} shows the neural network's average performance over the lossless images in LIVE1 dataset. For Q = 10, 30, 50, the model improves the PSNR (IPSNR) by 1.45 dB, 1.41 dB, 1.28 dB, and improves the SSIM (ISSIM) by 2.50\%, 1.16\%, 0.79\%, respectively. A performance comparison with existing machine learning models is shown in Table \ref {table: model comparison}. Here we emphasize that our proposed HR-CNN is a true end-to-end neural network, while all the reference models require additional pre-processing to convert k-space JPEG code to pixel space. The HR-CNN achieves leading performance under PSNR metrics. Its parameters are considerably fewer than that of models at comparable performance. Benefitting from the end-to-end fully convolutional structure, HR-CNN also demonstrates obvious advantages in efficiency. Its total number of trainable parameters is considerably less than that of existing models with comparable reconstruction quality. Numerical experiments also show that HR-CNN consumes 6.3s to decode and recover the LIVE1 datasets with GPU acceleration. In comparison, decoding the same set of images following standard JPEG protocol costs 2.81s on CPU. Classic decoding itself is equivalent to 45\% of HR-CNN's total processing time.

It should also be noticed that the absolute increase in the PSNR or the SSIM is not sufficient to fully demonstrate the power of the model. This emerges from the fact that JPEG tends to maintain the lower frequency components and reduce the accuracy when storing the high frequency components. Even at any given $Q$ factor, the JPEG compression protocol does not discard the same amount of information for all instances of digital images. For example, for an image consists of very complex geometric structures, the JPEG compression usually results in significant artifacts. This is because the original image is rich in high frequency components, whose accuracy are reduced after compression. On the contrary, for an image with no details, the JPEG protocol usually does not cause the same noticeable artifacts after compression. In this case, we should not expect a significant PSNR increase in the reconstructed image. The model's processing power in artifacts suppression needs to be best analyzed through both the average reported PSNR/ SSIM and their absolute increase through reconstruction.

\section{Conclusions}
A novel design of heterogeneous residual convolutional neural network is proposed for JPEG artifacts suppression. Directly starting from the JPEG code in k-space, we formalize the artifacts suppression task as an interactive process of decoding, macroblocking artifacts suppression and detail sharpness enhancement. A spectral decomposition mechanism with coded mask convolutional kernels is designed to fully address the uncorrelated nature of different spectral channels in the JPEG code. After expanding the throughput in each channel, the spectral snapshots are decoded heterogeneously to obtain the pixel space image. A residual learning segment then conceptually takes the responsibility for further detail sharpness enhancement. The fully convolutional nature and the elegant structure of the proposed network makes it relatively easy to train, and computationally efficient for online usage. Numerical results also validate that the proposed neural work demonstrates outstanding performance for multiple quantization factors.

\bibliographystyle{IEEEtran}
\bibliography{IEEEabrv,DLJPEG}
\nocite{*}

\end{document}